\documentclass[twocolumn,showpacs,amsfonts,aps,prc,nofootinbib,floatfix,%
superscriptaddress]{revtex4}

\usepackage{amsmath}
\usepackage{bm}
\usepackage{graphicx}

\newcommand{\VC}{{\tt VISHNU}}
\newcommand{\U}{{\tt UrQMD}}

\newcommand{\ecc}{\varepsilon}
\newcommand{\dNdy}{dN_\mathrm{ch}/dy}
\newcommand{\Npart}{N_\mathrm{part}}

\newcommand{\be}[1]{\begin{equation}\label{#1}}
\newcommand{\ee}{\end{equation}}

\newcommand{\vtwopt}{$v_2(p_T)$}
\newcommand{\eq}{{\,=\,}}

\def\La{\langle}
\def\Ra{\rangle}

\usepackage{color}

\begin{document}


\title{Hadron spectra and elliptic flow for $\bm{200\,A}$\,GeV Au+Au
       collisions from viscous hydrodynamics coupled to a Boltzmann cascade}
\date{\today}

\author{Huichao Song}
\affiliation{Nuclear Science Division,
             Lawrence Berkeley National Laboratory, Berkeley,
             California 94720, USA}

\author{Steffen A. Bass}
\affiliation{Department of Physics, Duke University, Durham,
             North Carolina 27708, USA}

\author{Ulrich Heinz}
\affiliation{Department of Physics, The Ohio State University,
             Columbus, Ohio 43210, USA}

\author{Tetsufumi Hirano}
\affiliation{Department of Physics, The University of Tokyo, Tokyo 113-0033,
             Japan}
\affiliation{Nuclear Science Division,
             Lawrence Berkeley National Laboratory, Berkeley,
             California 94720, USA}

\author{Chun Shen}
\affiliation{Department of Physics, The Ohio State University,
             Columbus, Ohio 43210, USA}

\begin{abstract}
It is shown that the recently developed hybrid code \VC, which couples
a relativistic viscous fluid dynamical description of the quark-gluon
plasma (QGP) with a microscopic Boltzmann cascade for the late
hadronic rescattering stage, yields an excellent description of
charged and identified hadron spectra and elliptic flow measured in
200\,$A$\,GeV Au+Au collisions at the Relativistic Heavy-Ion Collider
(RHIC). Using initial conditions that incorporate event-by-event
fluctuations in the initial shape and orientation of the collision
fireball and values $\eta/s$ for the specific shear viscosity of the
quark-gluon plasma that were recently extracted from the measured centrality
dependence of the eccentricity-scaled, $p_T$-integrated charged hadron
elliptic flow $v_{2,\mathrm{ch}}/\ecc$, we obtain universally good agreement
between theory and experiment for the $p_T$-spectra and differential
elliptic flow \vtwopt\ for both pions and protons at all collision
centralities.
\end{abstract}

\maketitle


\section{Introduction}
\label{sec1}

In a recent article \cite{Song:2010mg} we extracted the shear viscosity
to entropy density ratio $(\eta/s)_\mathrm{QGP}$ of the quark-gluon plasma
(QGP) created in heavy-ion collisions at RHIC by comparing experimental
data for the eccentricity-scaled elliptic flow $v_2/\ecc$ with calculations
performed with \VC\ \cite{Song:2010aq}, a hybrid model that describes
the QGP stage of the expansion of the collision fireball macroscopically
with viscous hydrodynamics (in which $(\eta/s)_\mathrm{QGP}$ enters as an
input parameter) but switches to a microscopic description in the late
hadronic phase where we solve the Boltzmann equation with \U\
\cite{Bass:1998ca}. An important step in this analysis was to ensure
that, as we compared theoretical curves for different $(\eta/s)_\mathrm{QGP}$
values with the experimental data in order to find the value preferred
by Nature, we maintained a good description of the total charged hadron
multiplicity and the hadron transverse momentum spectra as a function
of collision centrality. That this was indeed achieved was announced
in Ref.~\cite{Song:2010mg} and will be documented in this companion article.
We then proceed to demonstrate that, with the QGP shear viscosity extracted
in \cite{Song:2010mg}, \VC\ provides a good description of all
single-particle aspects of soft hadron production in 200\,$A$\,GeV Au+Au
collisions for which accurate measurements exist, over the entire
range of collision centralities.

\section{Methodology}
\label{sec2}

The various components of the viscous hydrodynamics+Boltzmann hybrid code
\VC\ have been described in Ref.~\cite{Song:2010aq} (see also
\cite{Bass:1998ca,Song:2007fn,Shen:2010uy}) to which we refer the reader 
interested in technical details. For the QGP fluid we approximate $\eta/s$ 
in the temperature range $T_\mathrm{c}{\,<\,}T{\,\alt\,}2T_\mathrm{c}$ by a 
constant \cite{Csernai:2006zz}. We switch from a hydrodynamic description of 
the QGP to the microscopic hadronic rescattering code {\tt UrQMD} at
temperature $T_\mathrm{sw}\eq165$\,MeV, adjusted to reproduce the chemical
freeze-out temperature measured in RHIC collisions
\cite{BraunMunzinger:2001ip}; as shown in \cite{Song:2010aq}, this is
at the same time the {\em highest} $T$ for which we have a valid microscopic
description and the {\em lowest} $T$ for which the macroscopic hydrodynamic
approach can be trusted.

As shown in Ref.~\cite{Song:2010mg,Romatschke:2007mq}, the QGP shear
viscosity $(\eta/s)_\mathrm{QGP}$ extracted from the experimentally
measured elliptic flow depends on the initial fireball eccentricity
$\varepsilon_\mathrm{part}\eq\frac{\La y^2{-}x^2 \Ra}{\La y^2{+}x^2 \Ra}$
where $x$ and $y$ label the coordinates along the short and long major axes
of the fireball in the plane transverse to the beam direction. (This
definition of $x$, together with the beam direction $z$, define the
``participant plane'', reflected in the subscript.) With
presently available tools this initial eccentricity cannot be directly
measured, and theoretically we have limited control over it. We here use
initial entropy density profiles from two popular geometric models
for the initial particle production in high-energy heavy-ion collisions,
the Monte Carlo Glauber model (MC-Glauber \cite{Miller:2007ri}), in
a version \cite{Hirano:2009ah} that uses finite size nucleons, and
the Monte Carlo fKLN (MC-KLN) model \cite{Hirano:2009ah,Drescher:2006ca,%
Kharzeev:2000ph}. These models give initial eccentricities that
differ (depending on centrality) by up to 25\% which we hope to
cover the physically reasonable range of uncertainty. We showed in
\cite{Song:2010mg} that this uncertainty in the initial eccentricity 
completely dominates the present error range in the phenomenological 
extraction of $(\eta/s)_\mathrm{QGP}$, and that future improvements in 
the accuracy of the experimentally extracted value of 
$(\eta/s)_\mathrm{QGP}$ cannot be achieved without obtaining better 
(experimental and/or theoretical) control over the initial fireball 
eccentricity.

%
\begin{figure*}[t]
\includegraphics[width=0.95\linewidth,clip=]{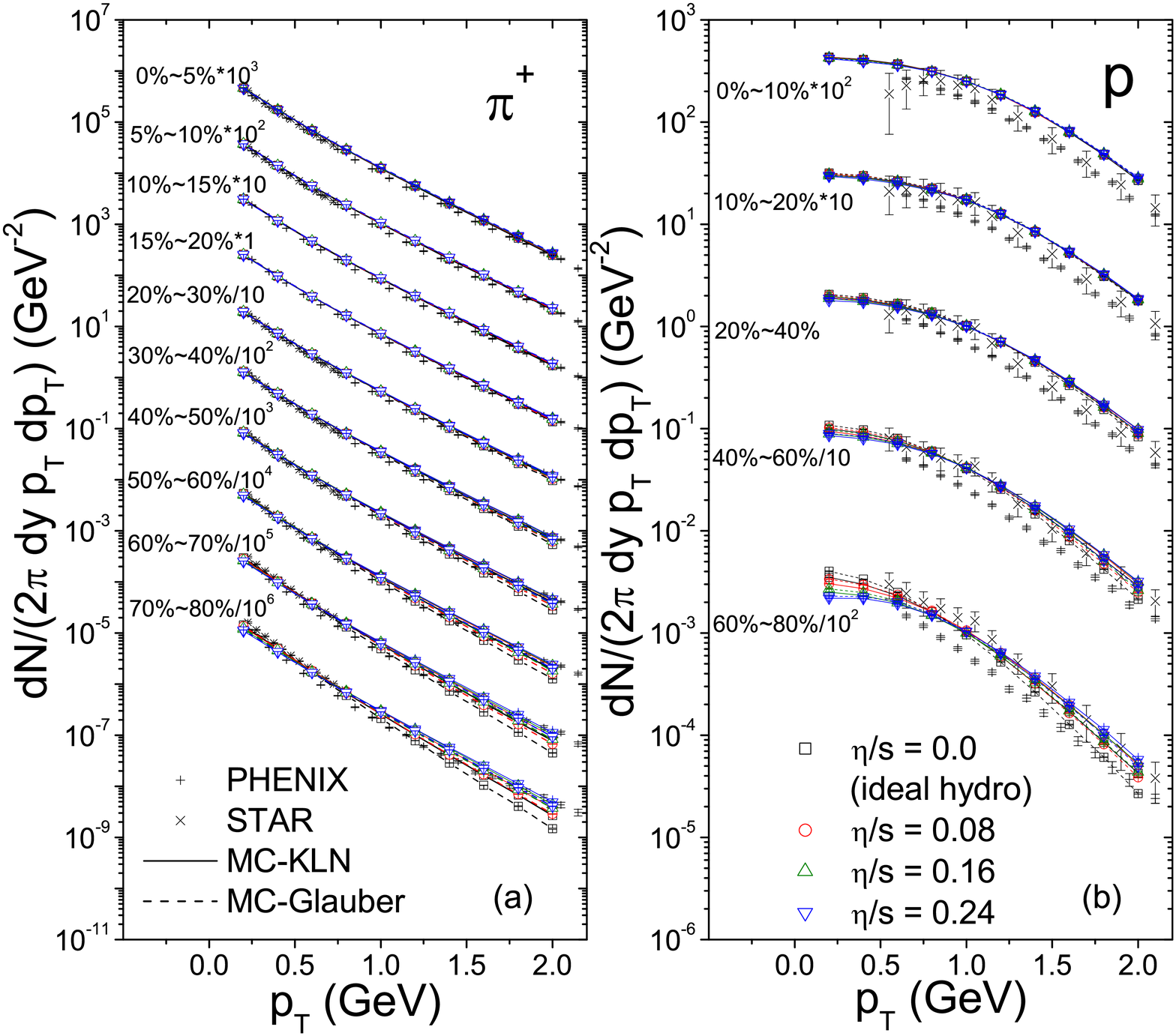}
\caption{\label{F1} (Color online)
$p_T$-spectra of pions (left) and protons (right) for 200\,$A$\,GeV Au+Au
collisions of different centralities as indicated. Data from the STAR
($\times$,\cite{Adams:2003xp,STARdata,:2008ez}) and PHENIX ($+$, 
\cite{Adler:2003cb}) experiments are compared with {\tt VISHNU} 
calculations using MC-Glauber (dashed lines) or MC-KLN initial 
conditions (solid lines) and different values $\eta/s$ for the QGP 
shear viscosity as indicated. Different $\eta/s$ values are associated 
with different starting times $\tau_0$ for the hydrodynamic evolution 
as discussed in the text. The STAR and PHENIX proton data shown in
the right column are feeddown-corrected by removing protons
from weak hyperon decays \cite{Adams:2003xp,Adler:2003cb}. Where necessary,
PHENIX yields from neighboring narrower centrality bins were averaged to 
obtain data in the wider centrality bins used by the STAR Collaboration. 
}
\end{figure*}
%

%
\begin{table*}[ht!] 
\caption{Mean ($\bar{\ecc}_\mathrm{part}{\,\approx\,}\La\ecc_\mathrm{part}\Ra$)
  and rms $\left(\ecc\{2\}\eq\sqrt{\La\ecc^2_\mathrm{part}\Ra}\right)$ 
  participant eccentricities, as well as $\La\ecc^\alpha_\mathrm{part}
  \Ra^{1/\alpha}$, with $\alpha$ computed from the event-plane resolution 
  $R$ as described in \cite{Ollitrault:2009ie,Art}, for the STAR 
  \cite{Adams:2004bi} and PHENIX \cite{Adare:2010ux} Au+Au experiments at 
  different collision centralities. For centralities between 10 and 50\%,
  PHENIX \cite{Adare:2010ux} published data in 5\% centrality increments. 
  For the purpose of comparing the PHENIX and STAR data we combined the 
  PHENIX data \cite{Adare:2010ux} from neighboring centrality bins, by 
  averaging the event-plane resolutions $R$ and the exponents $\alpha$ 
  corresponding to the two sub-bins when computing $\La\ecc^\alpha_
  \mathrm{part}\Ra^{1/\alpha}$ for the larger combined bin. All 
  eccentricities are calculated with the entropy density as weight.
  \label{T1}}
\label{tab:1} 
\begin{center}
\begin{ruledtabular}
\begin{tabular}{ccccccccccccc}
  centrality 
  &\vline &
  model &
  $\bar{\ecc}_\mathrm{part}$ &  
  $\sqrt{\La\ecc^2_\mathrm{part}\Ra}$ 
  &\vline &
  \begin{tabular}{c}
  \phantom{n} \\ $R$ \protect{\cite{Adams:2004bi}}
  \end{tabular} 
  &
  \begin{tabular}{c}
  STAR \\ $\alpha$ \protect{\cite{Ollitrault:2009ie,Art}}
  \end{tabular} 
  &
  \begin{tabular}{c}
  \phantom{n} \\ $\La\ecc^\alpha_\mathrm{part}\Ra^{1/\alpha}$
  \end{tabular} 
  &\vline &
  \begin{tabular}{c}
  \phantom{n}\\ $R$ \protect{\cite{Adare:2010ux}}
  \end{tabular} 
  &
  \begin{tabular}{c}
  PHENIX \\ $\alpha$ \protect{\cite{Ollitrault:2009ie,Art}}
  \end{tabular} 
  &
  \begin{tabular}{c}
  \phantom{n} \\ $\La\ecc^\alpha_\mathrm{part}\Ra^{1/\alpha}_{\phantom{\mathrm{R_{R_R}}}}$
  \end{tabular} 
  \\
  \hline\hline
  0--5\%   
  &\vline &
  \begin{tabular}{c}
  MC-Glauber \\ MC-KLN
  \end{tabular} 
  &
  \begin{tabular}{c}
  0.089 \\ 0.097
  \end{tabular} 
  &
  \begin{tabular}{c}
  0.101 \\ 0.109
  \end{tabular} 
  &\vline &
  0.61 & 1.52 &
  \begin{tabular}{c}
  0.095\\0.103
  \end{tabular}
  &\vline &
  0.51 & 1.66 &
  \begin{tabular}{c}
  0.097\\0.105
  \end{tabular} \\
  \hline
  5--10\%   
  &\vline &
  \begin{tabular}{c}
  MC-Glauber \\ MC-KLN
  \end{tabular} 
  &
  \begin{tabular}{c}
  0.139 \\ 0.172
  \end{tabular} 
  &
  \begin{tabular}{c}
  0.153 \\ 0.183
  \end{tabular} 
  &\vline &
  0.735 & 1.31 &
  \begin{tabular}{c}
  0.144\\0.175
  \end{tabular}
  &\vline &
  0.63 & 1.49 &
  \begin{tabular}{c}
  0.146\\0.178
  \end{tabular} \\
  \hline
  10--20\%   
  &\vline &
  \begin{tabular}{c}
  MC-Glauber \\ MC-KLN
  \end{tabular} 
  &
  \begin{tabular}{c}
  0.215 \\ 0.265
  \end{tabular} 
  &
  \begin{tabular}{c}
  0.230 \\ 0.277
  \end{tabular} 
  &\vline &
  0.816 & 1.18 &
  \begin{tabular}{c}
  0.218\\0.267
  \end{tabular}
  &\vline &
  0.720 & 1.33 &
  \begin{tabular}{c}
  0.220\\0.269
  \end{tabular} \\
  \hline
  20--30\%   
  &\vline &
  \begin{tabular}{c}
  MC-Glauber \\ MC-KLN
  \end{tabular} 
  &
  \begin{tabular}{c}
  0.299 \\ 0.360
  \end{tabular} 
  &
  \begin{tabular}{c}
  0.311 \\ 0.372
  \end{tabular} 
  &\vline &
  0.843 & 1.14 &
  \begin{tabular}{c}
  0.298\\0.362
  \end{tabular}
  &\vline &
  0.743 & 1.30 &
  \begin{tabular}{c}
  0.301\\0.364
  \end{tabular} \\
  \hline
  30--40\%   
  &\vline &
  \begin{tabular}{c}
  MC-Glauber \\ MC-KLN
  \end{tabular} 
  &
  \begin{tabular}{c}
  0.361 \\ 0.434
  \end{tabular} 
  &
  \begin{tabular}{c}
  0.378 \\ 0.447
  \end{tabular} 
  &\vline &
  0.825 & 1.16 &
  \begin{tabular}{c}
  0.364\\0.436
  \end{tabular}
  &\vline &
  0.704 & 1.36 &
  \begin{tabular}{c}
  0.367\\0.439
  \end{tabular} \\
  \hline
  40--50\%   
  &\vline &
  \begin{tabular}{c}
  MC-Glauber \\ MC-KLN
  \end{tabular} 
  &
  \begin{tabular}{c}
  0.414 \\ 0.493
  \end{tabular} 
  &
  \begin{tabular}{c}
  0.433 \\ 0.509
  \end{tabular} 
  &\vline &
  0.771 & 1.25 &
  \begin{tabular}{c}
  0.419\\0.497
  \end{tabular}
  &\vline &
  0.617 & 1.50 &
  \begin{tabular}{c}
  0.424\\0.501
  \end{tabular} \\
  \hline
  50--60\%   
  &\vline &
  \begin{tabular}{c}
  MC-Glauber \\ MC-KLN
  \end{tabular} 
  &
  \begin{tabular}{c}
  0.458 \\ 0.541
  \end{tabular} 
  &
  \begin{tabular}{c}
  0.481 \\ 0.561
  \end{tabular} 
  &\vline &
  0.677 & 1.41 &
  \begin{tabular}{c}
  0.468\\0.549
  \end{tabular}
  &\vline &
  0.489 & 1.69 &
  \begin{tabular}{c}
  0.475\\0.555
  \end{tabular} \\
  \hline
  60--70\%   
  &\vline &
  \begin{tabular}{c}
  MC-Glauber \\ MC-KLN
  \end{tabular} 
  &
  \begin{tabular}{c}
  0.497 \\ 0.581
  \end{tabular} 
  &
  \begin{tabular}{c}
  0.523 \\ 0.606
  \end{tabular} 
  &\vline &
  0.549 & 1.61 &
  \begin{tabular}{c}
  0.513\\0.597
  \end{tabular}
  &\vline &
  --- & --- & --- \\
  \hline
  70--80\%   
  &\vline &
  \begin{tabular}{c}
  MC-Glauber \\ MC-KLN
  \end{tabular} 
  &
  \begin{tabular}{c}
  0.528 \\ 0.621
  \end{tabular} 
  &
  \begin{tabular}{c}
  0.560 \\ 0.650
  \end{tabular} 
  &\vline &
  0.412 & 1.78 &
  \begin{tabular}{c}
  0.554\\0.645
  \end{tabular}
  &\vline &
  --- & --- & --- \\
\end{tabular}
\end{ruledtabular}
\end{center}
\end{table*}
%

Due to the finite number of nucleons colliding with each other in a
heavy-ion collision, the initial eccentricity of the density of secondary
particles produced in these collisions fluctuates from event to event,
as does the orientation of its major and minor axes relative to
the reaction plane \cite{Miller:2003kd} (defined by the directions of
the impact parameter and the beam). To account for these event-by-event
fluctuations on average, we use a Monte Carlo sampling procedure to
generate from the Glauber and fKLN models a large number of initial 
entropy density distributions whose shape and orientation fluctuate from 
event to event, recenter and rotate each distribution around the beam 
direction such that its short major axis $x$ aligns with the
direction of the impact parameter $\bm{b}$, sort them into centrality 
bins by $\Npart$ (the number of wounded nucleons), and then superimpose 
the distributions to obtain 
a smooth average density that has the correct average eccentricity for 
collisions in this centrality class.\footnote{\label{fn1}
  Strictly speaking, this procedure yields
  $\bar\ecc_\mathrm{part}{\,\equiv\,}\frac{\La y^2{-}x^2\Ra_{\bar{s}}}
  {\La y^2{+}x^2\Ra_{\bar{s}}}$ where $\La\dots\Ra_{\bar{s}}$ denotes
  the expectation value taken with the averaged entropy density obtained
  by superimposing many recentered and rotated Monte Carlo events; this
  is not identical with, but numerically very close to the ensemble-averaged
  participant eccentricity $\La\ecc_\mathrm{part}\Ra$ where for each
  event $\ecc_\mathrm{part}$ is computed as the analogous expectation
  value taken with the entropy density of that event.}
The elliptic flow resulting from the \VC\ evolution of this initial
profile is interpreted as the event-average $\La v_2\Ra$ for the selected
centrality class.

%
\begin{figure*}[t]
\includegraphics[width=\linewidth,clip=]{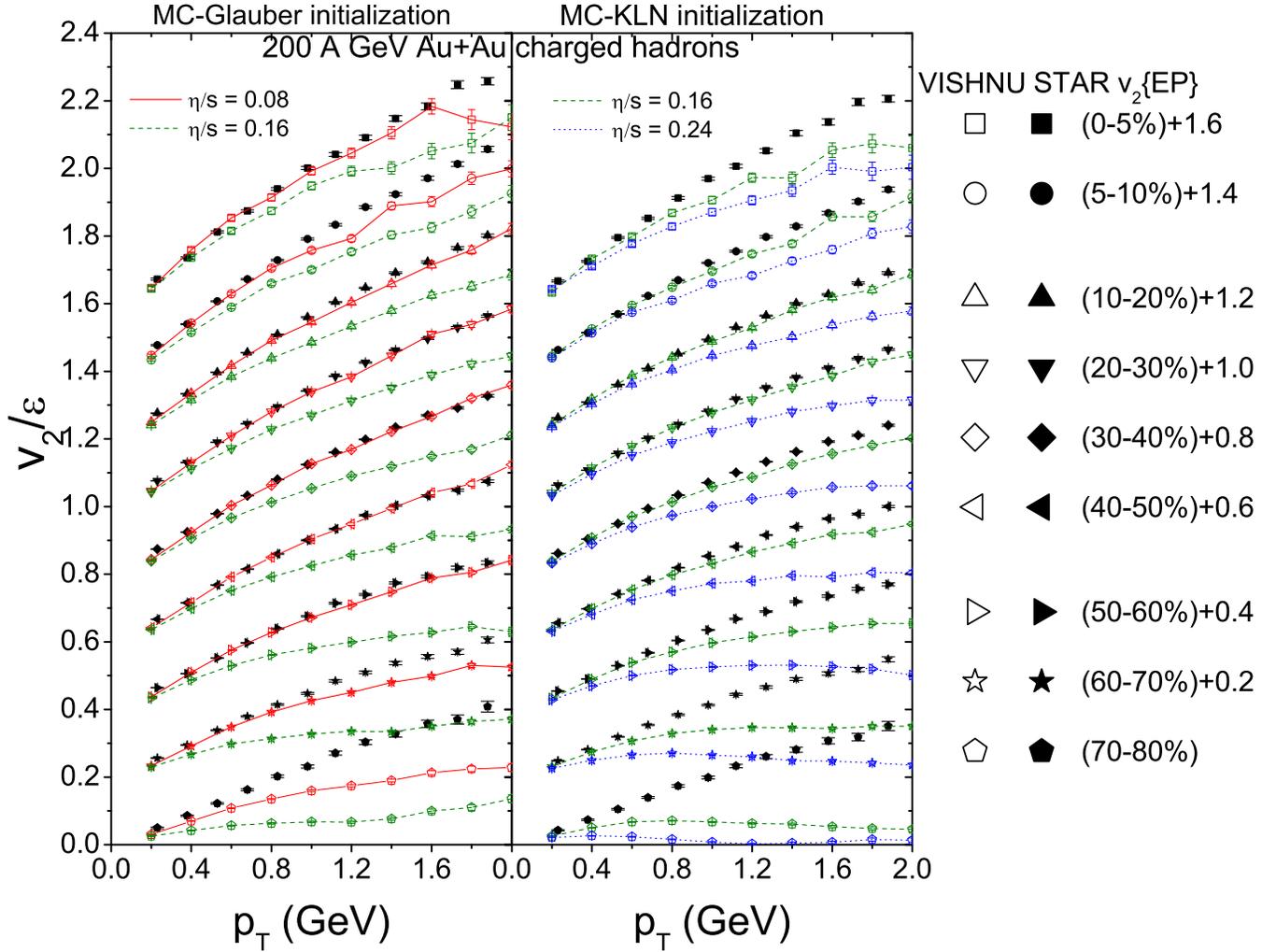}
\caption{\label{F2} (Color online)
Eccentricity-scaled elliptic flow $v_2/\ecc$ as function of $p_T$
for charged hadrons from 200\,$A$\,GeV Au+Au collisions at different
centralities. The experimental data (solid symbols) are $v_2\{\mathrm{EP}\}$ 
measurements from the STAR experiment \cite{Adams:2004bi}, scaled by 
$\langle\ecc_\mathrm{part}^\alpha\rangle^{1/\alpha}$ from the Monte 
Carlo Glauber model (left column) and the MC-KLN model (right column), 
respectively. $\alpha$ depends on the event-plane resolution $R$ 
and varies from one centrality bin to the next (see Table~\ref{T1}).  
Dashed and solid lines with open symbols are results from
\VC\ for two different values of $(\eta/s)_\mathrm{QGP}$ (0.08 and 0.16
for the MC-Glauber calculations, 0.16 and 0.24 for the MC-KLN calculations).
The theoretical lines show the ratio $\langle v_2\rangle/
\bar{\ecc}_\mathrm{part}$ where $\langle\dots\rangle$ denotes an average
over events, and $\bar{\ecc}_\mathrm{part}$ is the eccentricity of
the smooth average initial entropy density. Different symbols denote 
different collision centralities as indicated.
}
\end{figure*}
%

The ensemble-averaged initial entropy density is normalized such that,
after evolution with \VC, it reproduces the measured final charged hadron 
rapidity density $\dNdy$ in the most central collisions \cite{:2008ez}; 
due to viscous entropy production this is an iterative process, requiring 
two or three iterations. After normalization in central collisions, the 
centrality dependence of the initial entropy production is taken directly 
from the model (MC-Glauber or MC-KLN); for the MC-Glauber model we follow 
\cite{Hirano:2009ah,Hirano:2010jg} and assume a two-component (soft+hard) 
model with a small hard fraction ($\delta\eq0.14$ \cite{Hirano:2009ah}) for 
the entropy production. In \cite{Hirano:2009ah} this fraction was fixed 
within a hydro+Boltzmann hybrid approach using ideal fluid dynamics for 
the QGP; taking the same fraction in our viscous hydro+Boltzmann code 
ignores the centrality dependence of viscous heating. We have checked that 
its effects on the centrality dependence of the final $\dNdy$ are negligible 
relative to experimental uncertainties.

%
\begin{figure*}[t]
\includegraphics[width=\linewidth,clip=]{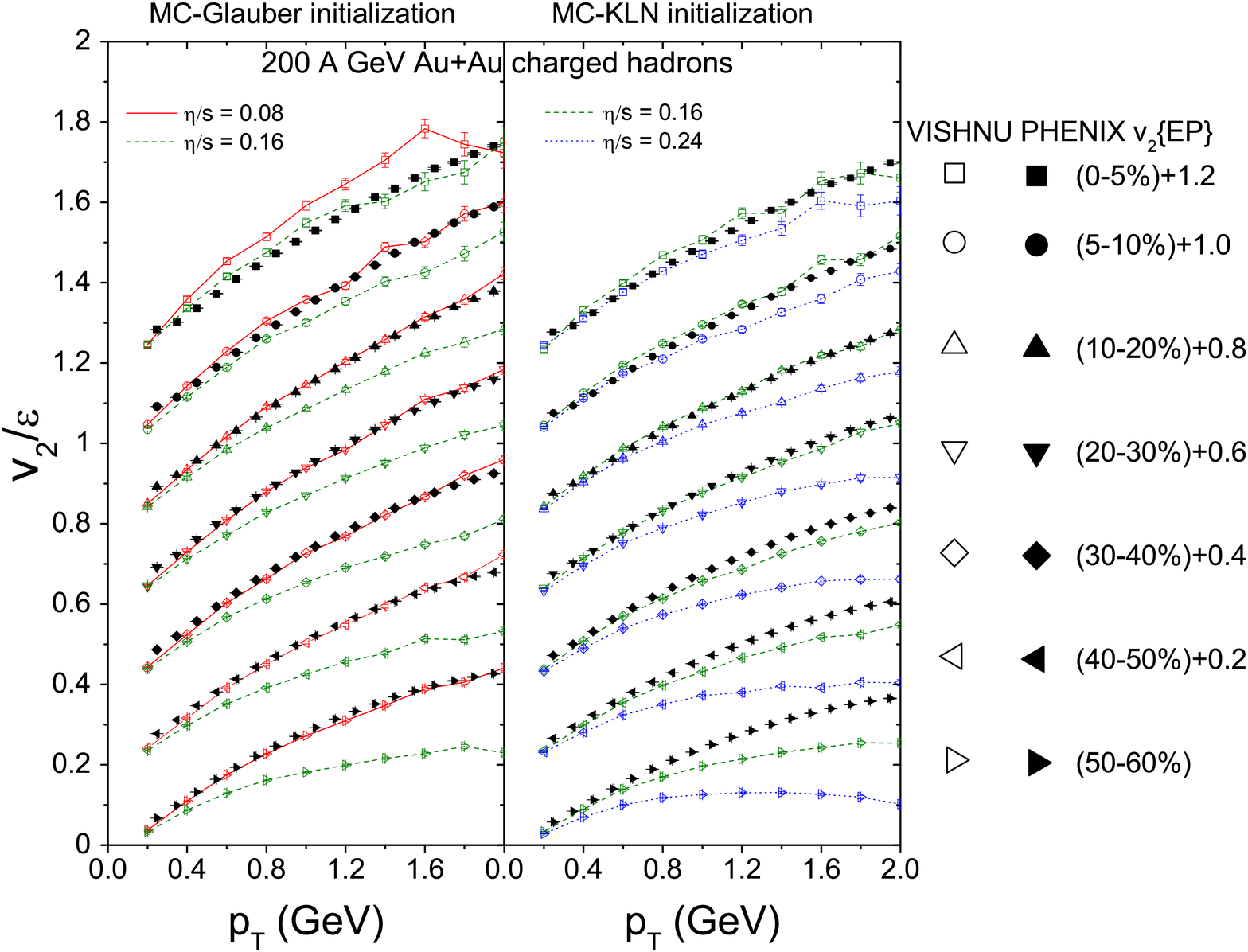}
\caption{\label{F3} (Color online)
Same as Fig.~\ref{F2}, but with $v_2\{\mathrm{EP}\}$ data from the
PHENIX Collaboration \cite{Adare:2010ux} instead of STAR data. See
Table~\ref{T1} for the effective eccentricities $\La\ecc^\alpha_
\mathrm{part}\Ra^{1/\alpha}$ used for each centrality bin. 
}
\end{figure*}
%

\section{Results: comparison of spectra and $\bm{v_2}$ to data}
\label{sec3}

Let us now begin discussing our results. All experimental data and
theoretical calculations are for 200\,$A$\,GeV Au+Au collisions.
Figure~\ref{F1} shows pion and proton transverse momentum spectra
from the STAR \cite{Adams:2003xp,STARdata,:2008ez} and PHENIX 
\cite{Adler:2003cb} Collaborations for the whole range of collision 
centralities, separated by multiplicative factors of 10 for clarity. The 
lines show \VC\ calculations for different values of $(\eta/s)_\mathrm{QGP}$
in the QGP phase, using either MC-KLN (solid) or MC-Glauber (dashed)
initial conditions. When changing $(\eta/s)_\mathrm{QGP}$ we have to
(i) renormalize the initial entropy density profiles by a
($b$-independent) constant factor to account for the change in viscous
entropy production, and (ii) adjust $\tau_0$ to account for the
additional radial acceleration caused by the transverse shear pressure
gradients. The latter increase with $(\eta/s)_\mathrm{QGP}$, leading to
more radial flow and flatter $p_T$-spectra unless we compensate by increasing
the starting time for the hydrodynamic evolution accordingly. The curves
shown in Fig.~\ref{F1} correspond to the following  parameter pairs
$(\eta/s,\,c\tau_0)$: (0,\,0.4\,fm), (0.08,\,0.6\,fm), (0.16,\,0.9\,fm),
and (0.24,\,1.2\,fm). We stop at $\eta/s\eq0.24$ since we will see that
larger QGP shear viscosities are excluded by the elliptic flow data.

Except for very peripheral collisions, the different lines in Fig.~\ref{F1}
overlap almost perfectly and thus are hard to distinguish optically.
This is intentional since it shows the approximate equivalence of the
different parameter pairs as far as the quality of the theoretical
description of the measured $p_T$-spectra goes. Differences between
theory and data are generally less than between data sets from the
different experiments. We note that the theoretical proton spectra are 
uniformly about 50\% larger than the PHENIX data but agree nicely with 
their slope; their normalization agrees somewhat better with the STAR 
data. Due to limited event statistics, \VC\ does not include protons from 
weak decays; in Fig.~\ref{F1}b we therefore compare with experimental data 
that have been corrected to eliminate feeddown protons. However, the 
feeddown correction methods used by PHENIX and STAR differ 
\cite{Adams:2003xp,Adler:2003cb}, and systematic uncertainties arising
from the feeddown correction are large. Keeping the differences between 
the experimental data sets in mind, \VC\ provides a very acceptable 
compromise description. We do note in passing that in the most peripheral 
bins a viscous treatment of the QGP appears to work better than treating 
it as an ideal fluid; assuming zero viscosity for the QGP gives too 
little radial flow and results in $p_T$ spectra for both pions and 
protons that are slightly too steep.

\begin{figure}[b!]
\includegraphics[width=\linewidth,clip=]{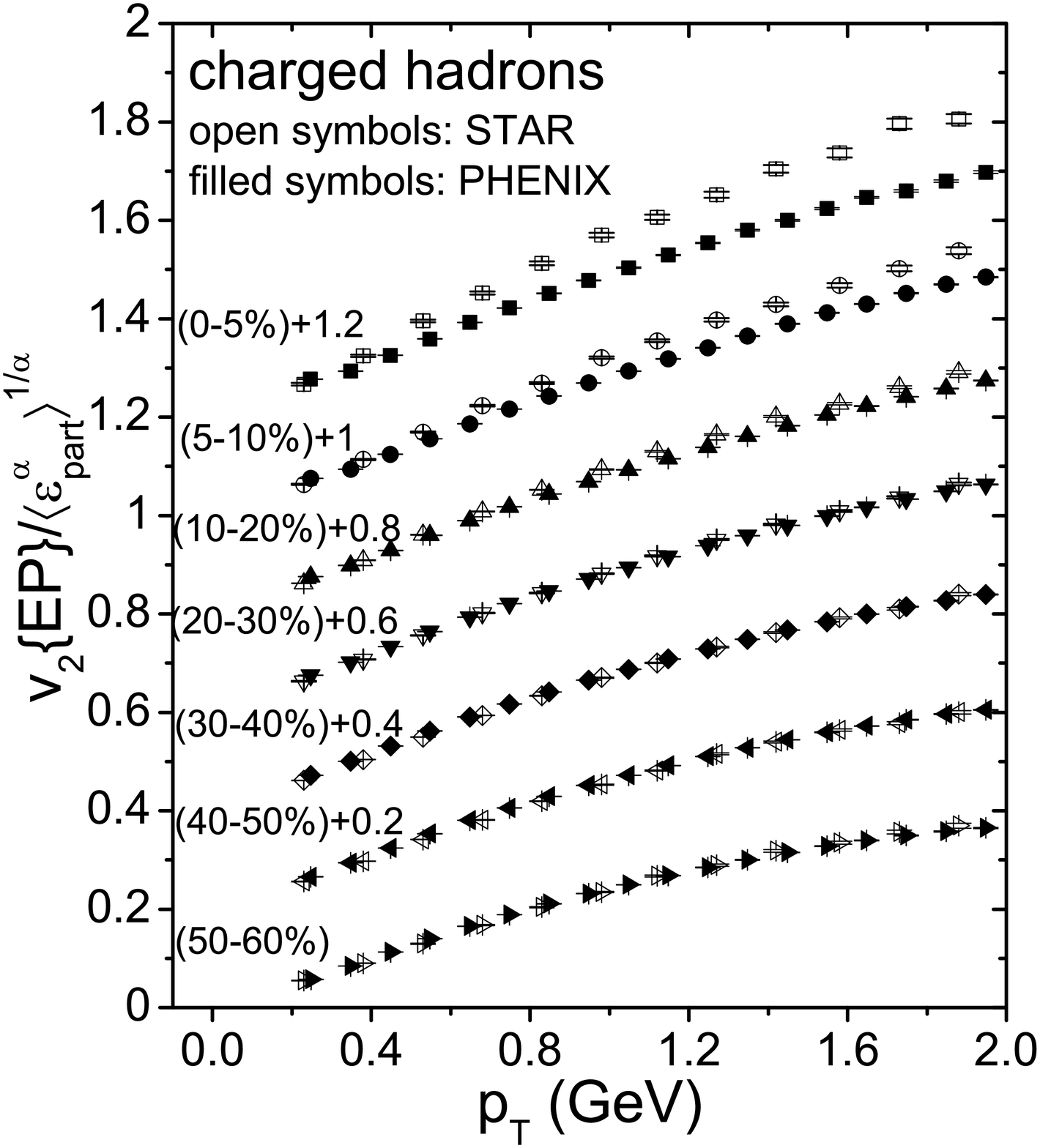}
\caption{\label{F4}
Comparison of the STAR \cite{Adams:2004bi} and PHENIX $v_2\{\mathrm{EP}\}$
data \cite{Adare:2010ux} used in Figs.~\ref{F2} and \ref{F3}. Both data 
sets are scaled by the effective eccentricity $\ecc{\,\equiv\,}\langle
\ecc_\mathrm{part}^\alpha\rangle^{1/\alpha}$, where $\alpha$ depends on 
the event-plane resolution $R$ \cite{Ollitrault:2009ie} and thus varies 
with centrality and from experiment to experiment (see Table~\ref{T1}).
}
\end{figure}

Figure~\ref{F2} shows the differential elliptic flow $v_2(p_T)$ for 
charged hadrons from Au+Au collisions at different centralities. 
Experimental data from the STAR Collaboration obtained with the 
event-plane method \cite{Adams:2004bi} are compared with \VC\ 
calculations for different QGP shear viscosities. $v_2\{\mathrm{EP}\}$ 
receives positive contributions from event-by-event flow fluctuations 
and non-flow effects \cite{Voloshin:2008dg}. The latter can be minimized 
by trying to decorrelate the determination of the event plane from the 
measurement of $v_2$, e.g. by employing a large rapidity gap between 
these measurements. Fluctuation effects can not be eliminated from the 
measurement, but both non-flow and fluctuations can be corrected for 
\cite{Ollitrault:2009ie}. In \cite{Song:2010mg} we used such corrected 
data to extract $(\eta/s)_\mathrm{QGP}$; in Figure~\ref{F2} we show the 
uncorrected $v_2\{\mathrm{EP}\}$ data directly as measured. To account 
for the fluctuation contribution we normalize them by
$\La\ecc_\mathrm{part}^\alpha\Ra^{1/\alpha}$ 
\cite{Ollitrault:2009ie,Alver:2008zza} where the exponent $\alpha$
depends on the experimental event-plane resolution $R$ and on details
of the $v_2$ extraction method \cite{Ollitrault:2009ie}. In Table~\ref{T1}
we have summarized for each centrality bin shown in Figs.~\ref{F2}-\ref{F4}
the event-plane resolution factors $R$ for the STAR and PHENIX experiments,
the corresponding $\alpha$ values obtained from the procedure described in 
Ref.~\cite{Ollitrault:2009ie}, as well as the corresponding values for
$\La\ecc_\mathrm{part}\Ra$, $\La\ecc_\mathrm{part}^2\Ra^{1/2}$, and
$\La\ecc_\mathrm{part}^\alpha\Ra^{1/\alpha}$. In Fig.~\ref{F2} we compare 
the experimental ratio $v_2\{\mathrm{EP}\}/\La\ecc_\mathrm{part}^\alpha
\Ra^{1/\alpha}$ with the theoretically calculated ratio 
$\La v_2\Ra/\bar{\ecc}_\mathrm{part}$. This is the correct comparison if 
$v_2\sim\ecc_\mathrm{part}$ event by event, as suggested by hydrodynamic 
simulations \cite{Alver:2010dn} (see, however, \cite{Qiu:2011iv}).

Figure~\ref{F2} demonstrates excellent agreement between \VC\ and the
experimental data, over the entire range of centralities except for the
two most peripheral bins, if we use $(\eta/s)_\mathrm{QGP}\eq0.08$ for
MC-Glauber (left column) and $(\eta/s)_\mathrm{QGP}\eq0.16$ for
MC-KLN initial conditions (right column).\footnote{Fig.~\ref{F2} also
  shows that in both cases the agreement is destroyed when increasing 
  $(\eta/s)_\mathrm{QGP}$ by $\frac{1}{4\pi}\eq0.08$.}
These values are a little smaller than, but consistent with the corresponding
values $(\eta/s)_\mathrm{QGP}{\,\simeq\,}0.1$ for MC-Glauber and
$(\eta/s)_\mathrm{QGP}{\,\simeq\,}0.2$ for MC-KLN that were extracted in
\cite{Song:2010mg} from the $p_T$-integrated, non-flow and fluctuation
corrected charged hadron $v_2$. Small non-flow effects in the
$v_2\{\mathrm{EP}\}$ data shown here, shifting them slightly upward, may
account for this difference.

To check this possibility, we show in Fig.~\ref{F3} the same comparison
with PHENIX data for $v_2\{\mathrm{EP}\}$ \cite{Adare:2010ux} where
the event plane was determined with counters several units of rapidity
away from the central region where $v_2$ was measured. The PHENIX data
should therefore be less affected by non-flow effects than the STAR data.
For centralities $>10\%$ the agreement between \VC\ calculations and
the PHENIX data is equally good as in Fig.~\ref{F2} for the STAR data,
with the same values for $(\eta/s)_\mathrm{QGP}$. In the two most central
bins, $5{-}10\%$ and $0{-}5\%$ respectively, the agreement deteriorates
significantly, with the PHENIX data pointing counterintuitively to
{\em larger} $(\eta/s)_\mathrm{QGP}$ values for central collisions than
in the other centrality bins. Figure~\ref{F4} shows that this results from
%
%
a disagreement between the two data sets in near-central collisions:
while the two data sets overlap excellently for centralities
$>20\%$, they increasingly diverge at small centralities, with a 30\%
difference between STAR an PHENIX in the $0-5\%$ centrality bin. It has
been pointed out that the excess of the STAR over the PHENIX data is
uniform in $p_T$ and could be explained by a 2\% shift in the centrality
definitions between the experiments \cite{private}. Where such a shift 
could arise from and which of the two definitions needs to be corrected 
is presently under study \cite{private2}. We conclude from Figs.~\ref{F2} 
-- \ref{F4} that (i) non-flow effects seem to be similar and likely small 
in both STAR and PHENIX $v_2\{\mathrm{EP}\}$ data for centralities between 
20\% and 60\%, (ii) if the STAR centrality definition is correct we have 
excellent agreement between \VC\ and the experimental charged hadron elliptic
flow $v_2(p_T)$ at all centralities, with $(\eta/s)_\mathrm{QGP}\eq0.08$
for MC-Glauber and $(\eta/s)_\mathrm{QGP}\eq0.16$ for MC-KLN initial
conditions, and (iii) if the PHENIX centrality definition is correct,
this uniform agreement is broken in the most central collisions for
which the PHENIX data appear to require larger effective
$(\eta/s)_\mathrm{QGP}$ values than at larger centralities.

%
\begin{figure*}[t]
\includegraphics[width=0.9\linewidth,clip=]{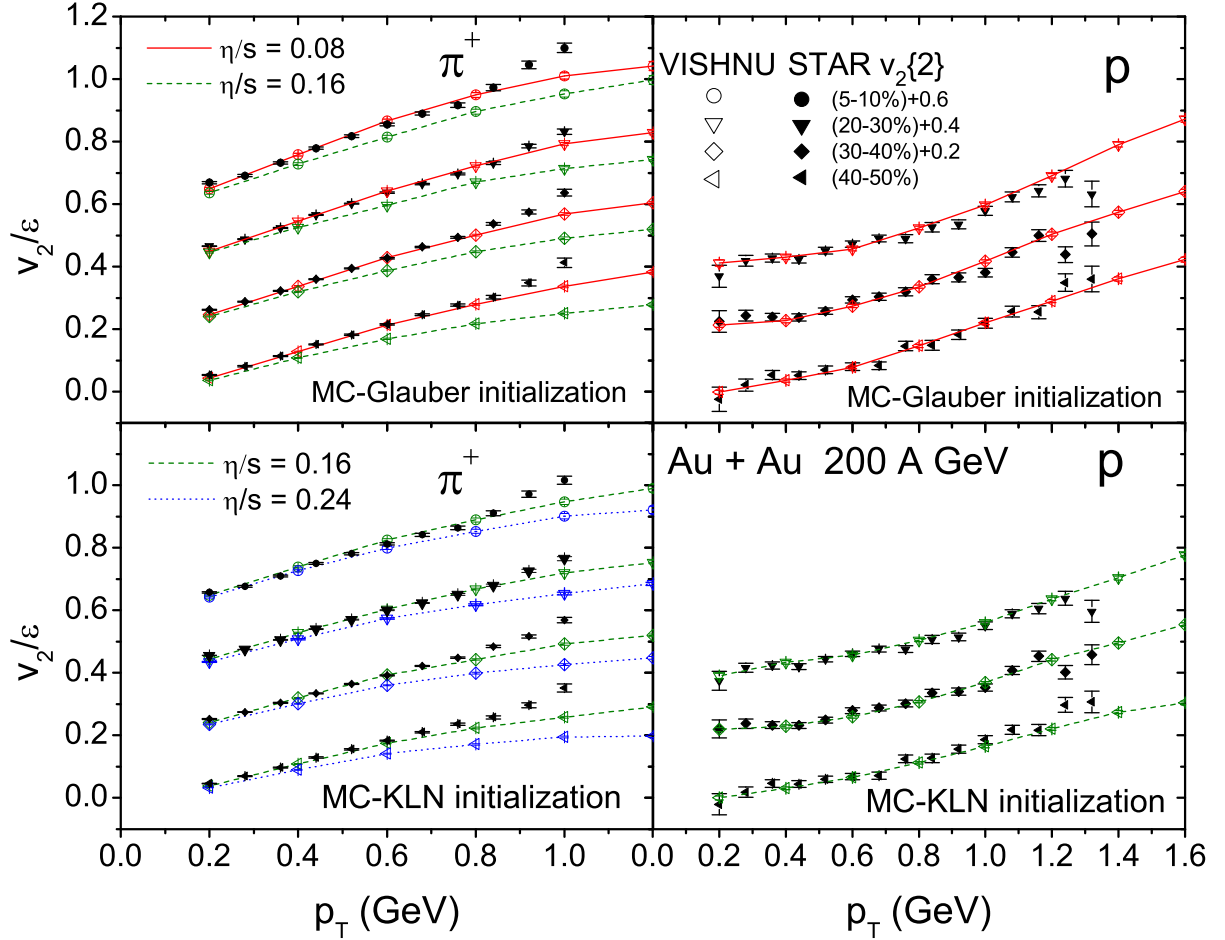}
\caption{\label{F5} (Color online)
Same as Fig.~\ref{F2}, but for identified pions (left column)
and protons (right column). Solid symbols denote measurements of
$v_2\{2\}/\langle\ecc_\mathrm{part}^2\rangle^{1/2}$ from the STAR
experiment \cite{Adams:2004bi}, solid and dashed lines with open
symbols show $\langle v_2\rangle/\bar{\ecc}_\mathrm{part}$
from \VC\ calculations with $(\eta/s)_\mathrm{QGP}\eq0.08$ and 0.16,
respectively, using MC-Glauber (top row) and MC-KLN (bottom row) initial
conditions. Different symbols denote different collision centralities
as indicated.
}
\end{figure*}
%

Figure~\ref{F5} demonstrates that the agreement of \VC\ using the
phenomenologically extracted $(\eta/s)_\mathrm{QGP}$ values from
Ref.~\cite{Song:2010mg} with the measured differential elliptic flow
carries over from all charged hadrons to identified pions and protons.
Data for protons that have sufficient statistical precision to
discriminate between different  $(\eta/s)_\mathrm{QGP}$ values exist
only for the mid-centrality range $20-50\%$. In very peripheral
collisions ($>60\%$ centrality) \VC\ has similar problems with
the pion $v_2(p_T)$ as we saw in Figs.~\ref{F2} and \ref{F3} for all
charged hadrons. (We comment on this discrepancy in the discussion in
Sec.~\ref{sec4}.) In the mid-centrality range Fig.~\ref{F5} shows excellent 
agreement between \VC\ with $(\eta/s)_\mathrm{QGP}\eq0.08$ for MC-Glauber 
and $(\eta/s)_\mathrm{QGP}\eq0.16$ for MC-KLN initial conditions and the
experimental data, in each of the three resp. four centrality bins shown.
The pion elliptic flow data in the left column reveal that for both
MC-Glauber and MC-KLN initial conditions this agreement breaks down if
$(\eta/s)_\mathrm{QGP}$ is increased by $\frac{1}{4\pi}\eq0.08$ above
the preferred value. For protons the calculation of $v_2(p_T)$ is
numerically costly (the elliptic flow signal and the number of protons per
event are both small), and we have therefore not done any calculations
for other than the preferred $(\eta/s)_\mathrm{QGP}$ values. However,
the proton data for these bins are precise enough that they would again
reject $(\eta/s)_\mathrm{QGP}$ values that differed by more than $1/4\pi$
from the values shown.

\section{Discussion}
\label{sec4}

The comparisons between theory and data discussed above prove that we can
extract the QGP shear viscosity from the centrality dependence of the
{\em $p_T$-integrated $v_2$} for charged hadrons and then use this value 
to obtain a very good overall description of the {\em $p_T$-differential 
$v_2$}. This works not only for the sum of all charged hadrons, but also 
for individual identified hadronic species, and it carries over from $v_2$ 
to their $p_T$-spectra which are nicely described over the entire range 
of collision centralities, except perhaps the most peripheral collisions. 

We do not recommend to try to extract $(\eta/s)_\mathrm{QGP}$ directly from 
the $p_T$-differential elliptic flow, for the following reasons. The main 
effect of shear viscosity that we exploit when extracting it from experiment 
is that it inhibits the hydrodynamic conversion of spatially anisotropic 
pressure gradients within the collision fireball into momentum anisotropies. 
As emphasized by Ollitrault \cite{Ollitrault:1992bk}, Heinz 
\cite{Heinz:2005zg}, and recently by Teaney \cite{Teaney:2009qa}, the 
$p_T$-integrated elliptic flow\footnote{More precisely: the 
  $p_T^2$-weighted elliptic flow $A_2{\,\equiv\,}\frac{\La p_x^2{-}p_y^2\Ra}
  {\La p_x^2{+}p_y^2\Ra}$, rather than $v_2\eq\La\cos(2\phi_p)\Ra\equiv
  \left\La\frac{p_x^2{-}p_y^2}{p_x^2{+}p_y^2}\right\Ra$.}
of the sum of all hadrons is the observable that has the most direct
relationship with the hydrodynamically generated total momentum
anisotropy.\footnote{Replacing ``all hadrons'' by ``all charged hadrons'' 
   is fine because of approximate symmetry between positive, negative and
   uncharged hadrons in ultrarelativistic collisions which generate almost
   baryon-free fireballs.} 
Hence it is {\em the total charged hadron $v_2$} that is controlled by 
$\eta/s$. How the hydrodynamically generated total momentum anisotropy is 
distributed among the different hadron species and in $p_T$ depends on the 
chemical composition and $p_T$ distributions of the hadrons 
\cite{Hirano:2005wx}. 

The correct theoretical description of the differential elliptic flow 
$v_2(p_T)$ of individual identified hadron species thus depends on the 
accurate reproduction of their yields and $p_T$ spectra which show much 
stronger sensitivities to details of the hydrodynamic simulation (such 
as initial conditions and shape of the initial density profiles) than 
the total momentum anisotropy itself. For example, in a purely hydrodynamic 
approach with Cooper-Frye freeze-out, even at extremely high collision 
energies where the total momentum anisotropy has time to fully saturate 
before freeze-out, lower freeze-out temperatures will lead to more radial 
flow; this affects the slope of the single-particle spectra, causing a 
concomitant change in the slope of $v_2(p_T)$ which is solely controlled 
by the fact that, after integration over $p_T$, the same total charged 
hadron elliptic flow must be reproduced as for a higher freeze-out 
temperature. In a hybrid approach such as ours, the Cooper-Frye procedure 
used to convert the hydrodynamic output into particle distributions 
involves a so-called ``$\delta f$ correction'' \cite{Teaney:2003kp} 
describing the deviation from local equilibrium on the conversion 
hypersurface; its form is presently not precisely known 
\cite{Dusling:2007gi,Monnai:2009ad}. For a given total charged hadron 
$v_2$, different parametrizations for $\delta f$ lead to different shapes 
of identified hadron spectra and $v_2(p_T)$. Bulk viscosity has very little 
effect on the total momentum anisotropy (and thus on the total charged 
hadron $v_2$) but affects the radial flow and hence the slopes of $p_T$ 
spectra and $v_2(p_T)$ \cite{Monnai:2009ad,Song:2009rh}. These 
interdependencies between the hadron $p_T$-spectra and their $p_T$-dependent 
elliptic flow make it hazardous to extract $(\eta/s)_\mathrm{QGP}$ from 
$v_2(p_T)$. With such an approach it is rather difficult to arrive at a 
uniformly good description of all soft hadron characteristics, and one 
easily ends up with different $(\eta/s)_\mathrm{QGP}$ values extracted from 
the elliptic flow of different hadron species or from collisions at different
centralities.

When using the $p_T$-integrated charged particle elliptic flow to
extract $(\eta/s)_\mathrm{QGP}$ one must, however, pay attention
to the fact that the measured elliptic flow fluctuates from event to event
and may be contaminated by non-flow contributions. This was emphasized
in Ref.~\cite{Song:2010mg} where we therefore used elliptic flow data that
had been corrected for non-flow effects and event-by-event fluctuations.
In the remainder of this article we elaborate on how the $p_T$-integrated
charged hadron elliptic flow $\La v_2\Ra$ from the dynamical model
\VC, calculated with the $(\eta/s)_\mathrm{QGP}$ values extracted in
\cite{Song:2010mg}, compares directly with various experimental
measurements that have not been corrected for fluctuation and non-flow
effects. The trends exposed in this comparison provide useful
insights.

Without the ability of doing event-by-event hydrodynamic simulations
\cite{Qiu:2011iv,Andrade:2008xh,Qin:2010pf}, we can at this moment 
account for event-by-event fluctuations of the initial fireball density 
distribution only on average, in one of two ways: Either we
recenter and rotate each Monte Carlo event, in order to align their
major and minor axes, before averaging the density distributions (this
produces an average density profile $\bar{s}_\mathrm{part}$ in the
``participant plane'', characterized by its average eccentricity
$\bar{\ecc}_\mathrm{part}{\,\approx\,}\La\ecc_\mathrm{part}\Ra$ (see
footnote \ref{fn1})), or we superimpose the densities without
recentering and rotating (producing a smooth average density profile
$\bar{s}_\mathrm{RP}$ in the ``reaction plane'', with ``standard''
eccentricity $\ecc_\mathrm{s}\eq\frac{\La y^2{-}x^2\Ra}{\La y^2{+}x^2\Ra}$
where (in contrast to footnote \ref{fn1}) the expectation values in
numerator and denominator are taken with $\bar{s}_\mathrm{RP}$). Both of
these methods incorporate (in different ways) the effect of
event-by-event fluctuations of the shape and orientation of the collision
fireball on the average initial eccentricity, but do not dynamically
propagate event-by-event fluctuations of the value of this eccentricity.
As a result, the hydrodynamic evolution produces a non-fluctuating
elliptic flow, and while the \U\ afterburner produces event-by-event
$v_2$ fluctuations, they are only due to finite number statistics and
not related to event-by-event fluctuations of the initial eccentricity
$\ecc$.

%
\begin{figure*}[hbt]
\includegraphics[width=0.85\linewidth,clip=]{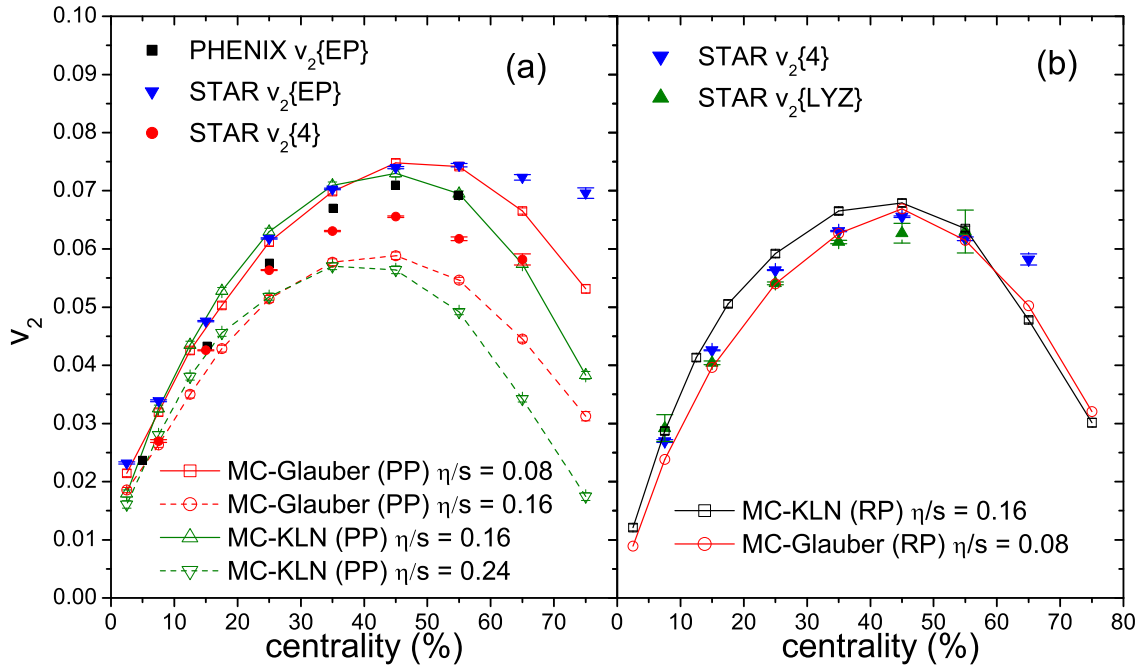}
\caption{\label{F6} (Color online)
Integrated charged hadron elliptic flow as a function of collision
centrality from the PHENIX \cite{Afanasiev:2009wq} and STAR
\cite{Adams:2004bi,Aamodt:2010pa} experiments are compared with
\VC\ calculations using participant plane (PP) averaged (a) and 
reaction plane (RP) averaged (b) initial conditions from the MC-KLN 
and MC-Glauber models and $(\eta/s)_\mathrm{QGP}$ values as indicated. 
In the STAR data and the calculations $v_2$ was integrated over the 
range 0.15\,GeV/$c<p_T<2$\,GeV/$c$; the PHENIX data were integrated over
0.2\,GeV/$c<p_T<8$\,GeV/$c$.
}
\end{figure*}
%

Various experimental techniques measure different varieties of $v_2$
which are affected in different ways by event-by-event flow fluctuations
(driven by event-by-event variations of the initial eccentricity) and
non-flow effects. Hydrodynamic simulations with smooth (non-fluctuating) 
initial conditions indicate a linear relationship $v_2\propto\ecc$ for 
not too large eccentricity \cite{Alver:2010dn}.\footnote{Event-by-event 
  hydrodynamic simulations with fluctuating initial conditions show that 
  flow anisotropy and eccentricity coefficients of different harmonic 
  order do not completely decouple from each other \cite{Qin:2010pf,%
  Qiu:2011iv}, and in both single-shot and event-by-event hydrodynamic 
  simulations we have seen evidence \cite{Qiu:2011iv} that in very central 
  collisions the (small) final elliptic flow is affected by several 
  harmonic eccentricity coeffients in the initial state (which, although 
  all small, are of similar order of magnitude), resulting in a nonlinear 
  dependence of $v_2$ on $\ecc$ for small $v_2$.}
If $v_2\propto\ecc$, the probability distribution of the final $v_2$ is 
directly related to that of the initial $\ecc$. For example, a measurement 
of $\sqrt{\La v_2^2\Ra}$ would yield values that are proportional to 
$\sqrt{\La\ecc^2\Ra}$, with the same proportionality constant as between 
$\La v_2\Ra$ and $\La\ecc\Ra$ \cite{Bhalerao:2006tp}. Unfortunately, 
quantities like $\ecc_\mathrm{part}\{2\}\eq\La\ecc_\mathrm{part}^2\Ra^{1/2}$ 
and $\ecc_\mathrm{part}\{4\}\eq\left[2\La\ecc_\mathrm{part}^2\Ra^2 -
\La\ecc_\mathrm{part}^4\Ra\right]^{1/4}$ that control the fluctuation 
contributions to $v_2\{2\}$ and $v_2\{4\}$ 
\cite{Voloshin:2008dg,Ollitrault:2009ie}, have a different centrality
dependence than the average eccentricities $\bar{\ecc}_\mathrm{part}$
and $\ecc_\mathrm{s}$ that characterize our hydrodynamic initial
conditions. For this reason it has been suggested in \cite{Bhalerao:2006tp}
to make comparisons between theory and experiment only with appropriately
normalized elliptic flows. In the absence of non-flow contributions
and for a linear mapping between $\ecc$ and $v_2$ in each event, one
has $\La v_2\Ra/\La\ecc_\mathrm{part}\Ra = v_2\{2\}/\ecc_\mathrm{part}\{2\}
 = v_2\{4\}/\ecc_\mathrm{part}\{4\}$ where the first ratio can be calculated 
in a single-shot hydrodynamic evolution of an average initial profile 
whereas the other two ratios can be measured if the initial eccentricity 
and its fluctuations are known from a model.

For Gaussian fluctuations, $P(\ecc_\mathrm{part}){\sim}\exp\bigl[-\frac
{(\ecc_\mathrm{part}{-}\bar{\ecc})^2}{2\sigma^2}\bigr]$, it is easy
to show \cite{Miller:2003kd} that $\ecc_\mathrm{part}\{2\}\eq\left[
\La\ecc_\mathrm{part}\Ra^2{+}\sigma^2\right]^{1/2}$ receives a positive
contribution from fluctuations whereas $\ecc_\mathrm{part}\{4\}\eq\left[
\La\ecc_\mathrm{part}\Ra^4-2\sigma^2\La\ecc_\mathrm{part}\Ra^2-\sigma^4
\right]^{1/4}$ is reduced relative to $\La\ecc_\mathrm{part}\Ra$. (In fact,
we can write in general $\ecc_\mathrm{part}\{4\}\eq\left[\La
\ecc_\mathrm{part}^2\Ra^2{-}(\La\ecc_\mathrm{part}^4\Ra -
\La\ecc_\mathrm{part}^2\Ra^2)\right]^{1/4}$ in terms of the difference between
two positive definite quantities which, for non-Gaussian distributions, can
become negative, in which case $\ecc_\mathrm{part}\{4\}$ is not defined.)
It has been observed in \cite{Voloshin:2007pc} that, for models where 
$\ecc_\mathrm{part}$ shows Bessel-Gaussian fluctuations 
\cite{Voloshin:2007pc,Voloshin:1994mz}, $\ecc_\mathrm{part}\{4\}$ agrees 
exactly with the reaction plane eccentricity 
$\La\ecc_\mathrm{RP}\Ra{\,\approx\,}\ecc_\mathrm{s}$,
and in \cite{Ollitrault:2009ie} that $v_2\{4\}$ is insensitive to
two-particle non-flow contributions. For these reasons, the authors
of \cite{Hirano:2010je} used hybrid model simulations with reaction-plane
averaged initial conditions for direct comparison with RHIC Au+Au and
recent LHC Pb+Pb data \cite{Aamodt:2010pa}. The validity of the assumption 
of (Bessel)-Gaussian eccentricity and flow fluctuations has been
challenged in \cite{Alver:2008zza} but was recently validaded for
the MC-Glauber and MC-KLN models for Au+Au collisions at centralities 
of up to about 40\% \cite{Qiu:2011iv}; for more peripheral collisions,
the assumption breaks down. We here compare results obtained from both 
participant-plane and reaction-plane averaged initial conditions with 
$p_T$-integrated $v_2\{4\}$ data.

In Fig.~\ref{F6} we compare \VC\ results with STAR and PHENIX data for
the integrated charged hadron elliptic flow as function of collision
centrality. The STAR and PHENIX data are integrated over slightly
different $p_T$ ranges; correcting the PHENIX data for the somewhat
smaller lower $p_T$ cutoff used by STAR and in the calculations would move
them slightly down. In the absence of non-flow contributions and the limit
of small fluctuations, $v_2\{\mathrm{EP}\}{\,\approx\,}\sqrt{\La v_2
\Ra^2{+}\sigma_v^2}$ and $v_2\{4\}{\,\approx\,}\sqrt{\La v_2
\Ra^2{-}\sigma_v^2}$. Non-flow effects would push $v_2\{\mathrm{EP}\}$
further up but leave $v_2\{4\}$ unchanged. Calculations with
reaction-plane averaged initial conditions of eccentricity
$\bar{\ecc}_\mathrm{part}$ (Fig.~\ref{F6}a) should thus fall
between $v_2\{\mathrm{EP}\}$ and $v_2\{4\}$, perhaps a bit
closer to $v_2\{4\}$ if the $v_2\{\mathrm{EP}\}$ data are affected
by non-flow. The STAR $v_2\{\mathrm{EP}\}$ data lie above those of
PHENIX, consistent with the expectation that the PHENIX data should
have less non-flow contributions (if any at all), but some of the
difference between the data set (especially at small centralities) may
also originate from a shift in the centrality definition \cite{private}.
Except for the most peripheral centralities, our calculations lie
above the PHENIX and roughly on the STAR $v_2\{\mathrm{EP}\}$ data
and overpredict the STAR $v_2\{4\}$ data. This indicates that the
chosen $(\eta/s)_\mathrm{QGP}$ values (0.08 for MC-Glauber and 0.16 for
MC-KLN initial conditions) are slightly too small (but not by much,
as seen by the fact that increasing $(\eta/s)_\mathrm{QGP}$ by 0.08
leads to a strong underprediction of all data sets), and that
the slightly larger values of 0.10 for MC-Glauber and 0.20 for
MC-KLN extracted in \cite{Song:2010mg} from fluctuation-corrected
$v_2$ data would give better agreement here, too. Fig.~\ref{F6}b
shows that with reaction-plane averaged initial conditions the \VC\
results agree very well with the $v_2\{4\}$ data, supporting the
argument \cite{Bhalerao:2006tp,Voloshin:2007pc} that $\La\ecc_\mathrm{RP}
\Ra{\,\approx\,}\ecc_\mathrm{s}$ provides a good substitute for
$\ecc_\mathrm{part}\{4\}$. Again, using the slightly larger
$(\eta/s)_\mathrm{QGP}$ values from \cite{Song:2010mg} would further
improve the agreement.

We note the inability of \VC\ to describe the elliptic flow in
the most peripheral collisions where the experimentally measured
values remain large whereas the theoretical predictions decrease rapidly
with increasing impact parameter. This drop is related to the decreasing
lifetime of the fireball (which is even shorter for runs with MC-KLN initial
conditions than for Glauber profiles, due to the sharper edges of
the MC-KLN profiles which lead to faster radial acceleration). Shorter
lifetimes leave less time for generating elliptic flow in the fluid dynamic
QGP stage, and the highly dissipative hadronic stage cannot compensate
for this. Calculations with an ideal hydro+cascade model that use a
different hadronic rescattering algorithm ({\tt JAM} instead of \U)
appear to share this feature \cite{Hirano:2010je}. The fact that
the $v_2\{4\}$ data do not show this decrease indicates that non-flow
contributions (which are not included in the model) are not to blame.
We do not know how to obtain larger $v_2$ values from the model at large
impact parameters.

\section{Summary and conclusions}
\label{sec5}

We have utilized the recently developed hybrid code \VC, which couples
a relativistic viscous fluid dynamical description of the quark-gluon
plasma (QGP) with a microscopic Boltzmann cascade for the late hadronic 
rescattering stage, to calculate charged and identified hadron spectra 
and elliptic flow measured in 200\,$A$\,GeV Au+Au collisions at the 
Relativistic Heavy-Ion Collider (RHIC). We find that, after suitable
readjustments of initial conditions, the $p_T$ spectra of identified
hadrons (pions and protons) are rather insensitive to the choice of the 
value of the specific shear viscosity $\eta/s$, whereas the 
eccentricity-scaled elliptic flow $v_2/\ecc$ shows strong sensitivity 
to $\eta/s$. Using initial conditions that incorporate event-by-event
fluctuations in the initial shape and orientation of the collision
fireball and values $(\eta/s)_\mathrm{QGP}$ for the specific shear 
viscosity of the quark-gluon plasma that were recently extracted from 
the measured centrality dependence of the eccentricity-scaled, 
$p_T$-integrated charged hadron elliptic flow $v_{2,\mathrm{ch}}/\ecc$ 
\cite{Song:2010mg}, we were able to obtain universally good agreement
between theory and experiment for the $p_T$-spectra and differential
elliptic flow \vtwopt\ for both pions and protons at all collision
centralities. Our analysis validates the constraints on $\eta/s$ reported 
in our previous work, namely that the QGP shear viscosity for
$T_\mathrm{c}{\,<\,}T{\,\alt\,}2T_\mathrm{c}$ lies within the range
$1<4\pi(\eta/s)_\mathrm{QGP}<2.5$, with the remaining uncertainty
dominated by insufficient theoretical control over the initial
source eccentricity $\varepsilon$.


\acknowledgments{We gratefully acknowledge fruitful discussions with
P. Huovinen, H. Masui, A. Poskanzer, S. Voloshin, and A. Tang. We 
specifically thank A. Poskanzer for computing for us the $\alpha$
values listed in Table~\ref{T1} and R. Snellings for providing some of 
the data shown in Fig.~\ref{F6}. This work was supported by the 
U.S.\ Department of Energy under Grants No. DE-AC02-05CH11231, 
DE-FG02-05ER41367, \rm{DE-SC0004286}, and (within the framework of the 
JET Collaboration) \rm{DE-SC0004104}. T.H. acknowledges
support through Grant-in-Aid for Scientific Research No. 22740151 and
through the Excellent Young Researchers Oversea Visit Program (No. 213383)
of the Japan Society for the Promotion of Science. We gratefully acknowledge
extensive computing resources provided to us by the Ohio Supercomputer Center.}

\appendix

\section{}

In this Appendix we add a few aspects that, due to space limitations,
were left out from the discussion in Ref.~\cite{Song:2010mg} of  the 
almost universal dependence of the eccentricity-scaled elliptic flow 
$v_2/\ecc$ on the charged hadron multiplicity density $(1/S)\dNdy$ on
which our extraction of the QGP shear viscosity $(\eta/s)_\mathrm{QGP}$
from RHIC data was based. Specifically, we show that the universality
of this dependence (i.e. the feature that it only depends on 
$(\eta/s)_\mathrm{QGP}$ but not on any details of the initial 
conditions for the hydrodynamic evolution which affect the initial
eccentricity $\ecc$ and transverse area $S$ of the expanding fireball)
holds not only for the participant-plane averaged fluctuating initial 
profiles used in \cite{Song:2010mg} but also for the reaction-plane
averaged profiles used here in Fig.~\ref{F6}b. Furthermore, it is 
insensitive to the smearing area $\sigma_s$ used in the MC-Glauber model
of Ref.~\cite{Hirano:2009ah} that describes the width of the transverse 
distribution of matter created in each nucleon-nucleon collision. However,
the source eccentriciy $\ecc$ itself depends on this smearing area, and
hence the $(\eta/s)_\mathrm{QGP}$ value extracted by comparing the 
universal theoretical $v_2/\ecc$ vs. $(1/S)\dNdy$ curves for different
$\eta/s$ with a given set of experimental $v_2$ vs. $\dNdy$ data also
depends on this parameter.  

%
\begin{figure}[b]
\includegraphics[width=\linewidth,clip=,angle=270]{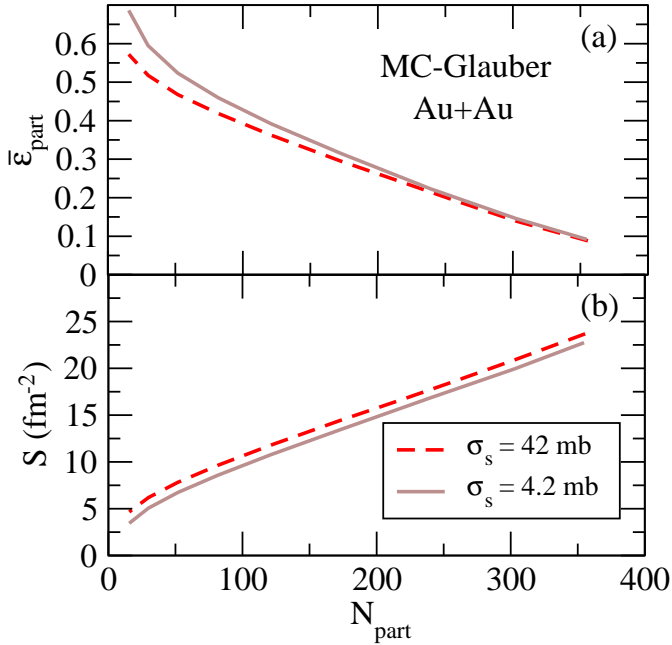}
\caption{\label{F7} (Color online)
Eccentricity $\bar{\ecc}_\mathrm{part}$ (a) and transverse area $S$ (b) of the 
particiant-plane averaged initial entropy density distribution from 
the MC-Glauber model, for smearing areas $\sigma_s\eq42$ and 4.2\,mb, 
respectively. (See also Fig.~4 in Ref.~\cite{Alver:2008zza} for 
comparison.)
}
\end{figure}
%

In the traditional MC-Glauber model, one samples the positions of nucleons
according to the nuclear density distributions of the two colliding nuclei
and calculates the participant eccentricity from the transverse positions
of the wounded nucleons and/or binary collision points, described by
$\delta$-functions in the transverse plane. Hirano and Nara 
\cite{Hirano:2009ah} pointed out that, since the measured nuclear density 
distribution represents a folding of the distribution of nucleon centers
with the finite size of each nucleon, the nuclear distribution used for 
sampling the positions of the nucleon centers must be described by different 
Woods-Saxon parameters than the measured nuclear density. Following
Ref.~\cite{Hirano:2009ah}, we therefore use for the distribution of nucleon 
centers in a Au nucleus a Woods-Saxon distribution with radius 
$R_\mathrm{Au}\eq6.42$\,fm and surface thickness
$d_\mathrm{Au}\eq0.44$\,fm (instead of the frequently used parameters 
$R_\mathrm{Au}^\mathrm{meas.}\eq6.38$\,fm and 
$d_\mathrm{Au}^\mathrm{meas.}\eq0.535$\,fm that describe the measured
nuclear density distribution of Au). We then distribute the entropy of 
particles emitted by a wounded nucleon or created in a binary nucleon-nucleon 
collision homogeneously in a cylinder of radius $r_s\eq\sqrt{\sigma_s/\pi}$ 
(where $\sigma_s$ is the so-called ``transverse smearing area''), centered 
at the position of the wounded nucleon or the collision point and aligned 
with the beam direction. (The same procedure was used in 
Ref.~\cite{Alver:2008zza} without, however, first correcting the Woods-Saxon
parameters of the distribution of nucleon centers for the finite nucleon 
size.)

The value of the smearing area $\sigma_s$ is not known {\em a priori} 
since it depends on unknown aspects of pre-thermal decoherence and entropy 
production processes. Theoretically, it is limited from below by the 
uncertainty principle which does not permit localization of the production 
points of secondary particles with average transverse momentum $\La p_T\Ra$ 
to an average distance $r_s{\,<\,}1/\La p_T\Ra$ from the classical collision 
point. We explore the choices $\sigma_s\eq42$\,mb \cite{Alver:2008zza,%
Hirano:2009ah} and 4.2\,mb. The smaller value is an approximation to 
pointlike secondary particle production (our code for calculating the 
participant-plane averaged initial density requires a non-vanishing 
$\sigma_s$); theoretically, it is disfavored by the above uncertainty 
argument.       

Figure~\ref{F7} shows the eccentricity (panel (a)) and transverse area 
(panel (b)) of the participant-plane averaged initial entropy density 
distribution as a function of the number $\Npart$ of participant 
(``wounded'') nucleons in Au+Au collisions, for two values of the smearing 
area $\sigma_s$. We see that a smaller smearing area (more pointlike 
particle production) leads to larger initial fireball eccentricities 
$\ecc_\mathrm{part}$ and smaller transverse areas $S$. While the effect 
of varying $\sigma_s$ on $S$ is simply an offset, for $\ecc_\mathrm{part}$ 
it leads to a change in the slope of its centrality dependence: changing 
$\sigma_s$ affects $\ecc_\mathrm{part}$ more strongly in peripheral 
collisions (where the nuclear overlap region is small and strongly 
deformed) than in central ones \cite{Alver:2008zza}.

%
\begin{figure}[h!]
\hspace*{3.5mm}
\includegraphics[height=6cm,clip=,angle=0]{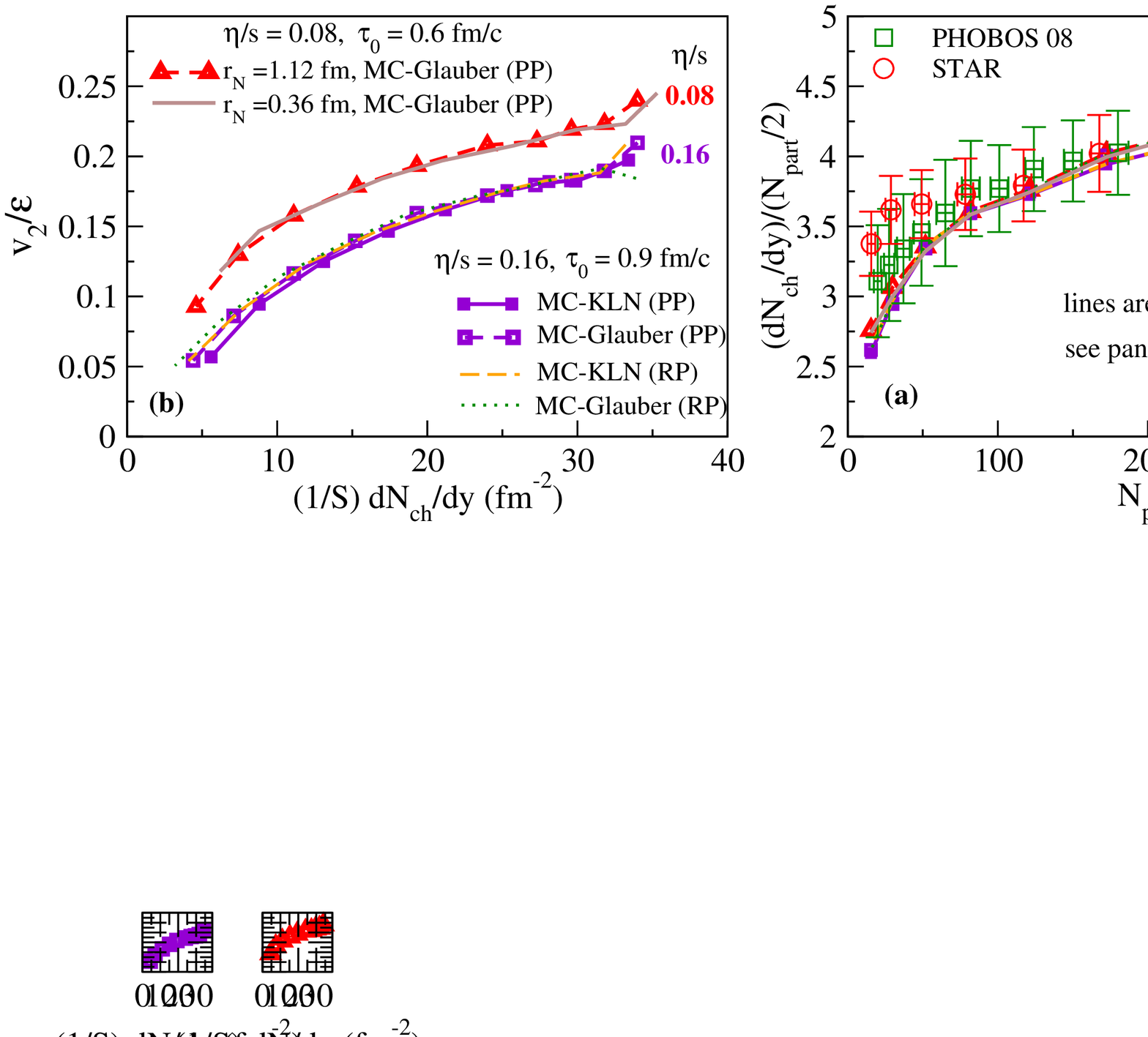}\\
\includegraphics[height=6cm,clip=,angle=0]{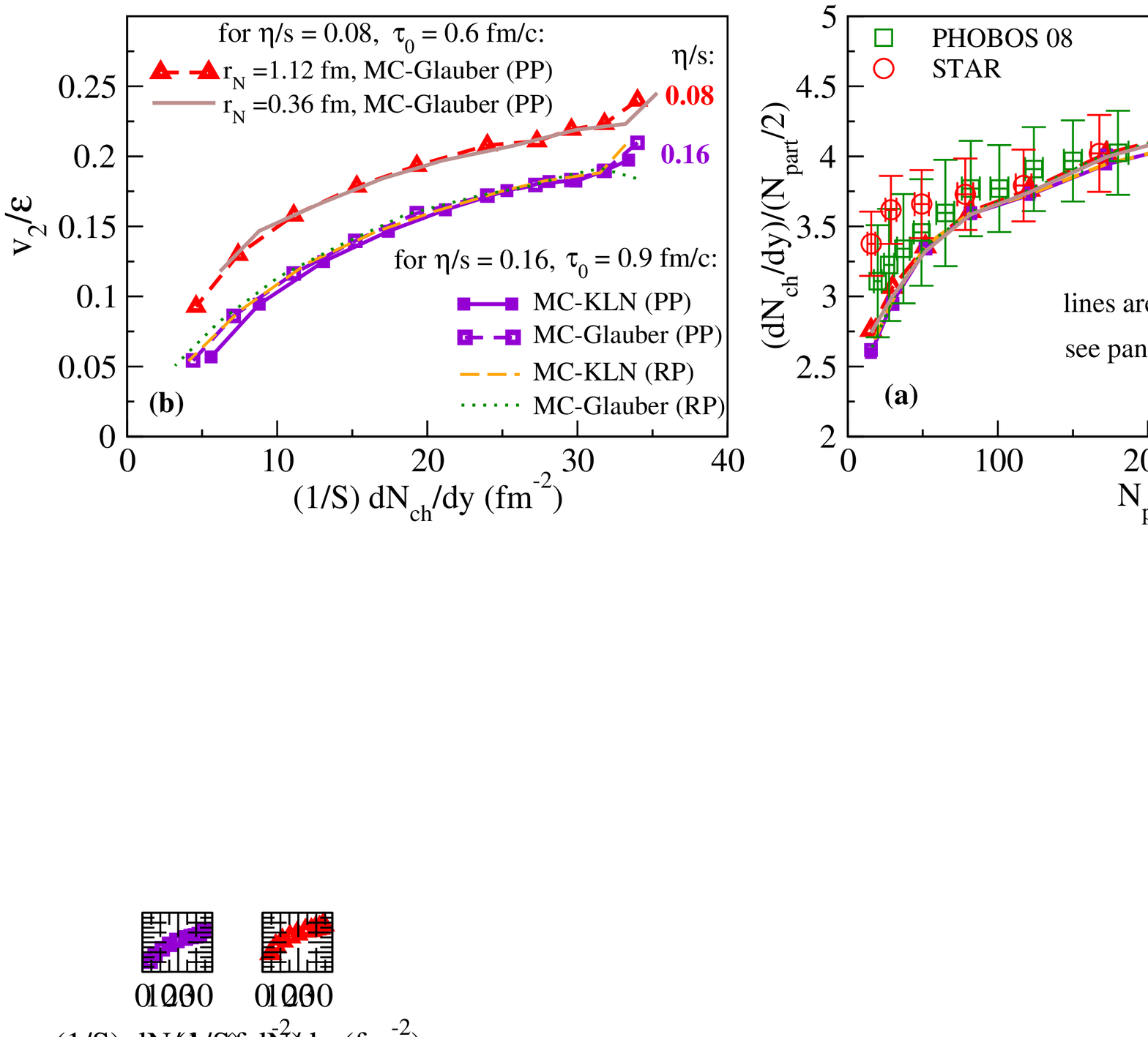}
\caption{\label{F8} (Color online) 
(a) Centrality dependence of the charged hadron rapidity density per
participant pair $(\dNdy)/(N_\mathrm{part}/2)$. Experimental data are
from STAR \cite{:2008ez} and PHOBOS \cite{Back:2004dy}, using 
$\dNdy\eq1.16\,dN_\mathrm{ch}/d\eta$ for the PHOBOS data. Theoretical
lines are explained in the text. 
(b) Eccentricity-scaled elliptic flow $v_2/\varepsilon$ as function of 
multiplicity density $(1/S)(\dNdy)$, for different values of 
$(\eta/s)_\mathrm{QGP}$, using participant-plane (PP) and reaction-plane 
(RP) averaged initial entropy density profiles from the MC-KLN and 
MC-Glauber models, normalized to $(\dNdy)_\mathrm{max}\eq810$ in the 
most central ($0{-}5\%$) Au+Au collisions. Insensitivity to the smearing 
area $\sigma_s$ in the MC-Glauber models is also shown.
}  
\end{figure}
%

Figure~\ref{F8} illustrates that changing the smearing area in the MC-Glauber
model has no effect on the centrality depenence of the produced charged
hadron multiplicity (panel (a)) nor on the universality of $v_2/\ecc$ vs. 
$(1/S)\dNdy$ (panel (b)). In Fig.~\ref{F8}b one sees (see solid brown line 
without symbols for $(\eta/s)_\mathrm{QGP}\eq0.08$) that, for smaller 
$\sigma_s\eq4.2$\,mb, the reduced fireball area shown in Fig.~\ref{F7}b 
shifts the entire curve towards the right. The shift is, however, not 
horizontal but rather diagonal such that, where they overlap, the shifted 
curve lies on top of the line for the larger value $\sigma_s\eq42$\,mb.
The upward component of the shift of the $\sigma_s\eq4.2$\,mb line arises 
from an increased QGP lifetime, due to the larger initial entropy density 
resulting from the smaller initial area $S$; a longer QGP lifetime in turn 
results in a larger momentum anisotropy at the beginning of the hadronic 
stage, since at RHIC energies the QGP never lives long enough for the fireball 
eccentricity to completely decay before hadronization. Larger momentum
anisotropy at the beginning of the hadronic rescattering stage leads to
more elliptic flow for the finally emitted hadrons.  

In Figure~\ref{F8} we also show curves obtained from \VC\ using reaction-plane
averaged (RP) initial profiles instead of participant-plane averaged (PP) 
ones. One sees that the centrality-dependence of the final charged 
multiplicity per participant (panel (a)) and the dependence of $v_2/\ecc$ 
on the multiplicity density $(1/S)\dNdy$ are insensitive to how we
average the fluctuating initial profiles from the MC-KLN and MC-Glauber
models when constructing the smooth initial entropy density profile
for the hydrodynamic evolution. Due to numerical cost we here show only 
curves for shear viscosity $(\eta/s)_\mathrm{QGP}\eq0.16$ for both 
MC-Glauber and MC-KLN models. We have made spot checks to convince 
ourselves that the scaling shown in Figure~\ref{F8} also works for 
other choices of $(\eta/s)_\mathrm{QGP}$.

\begin{figure}[h!]
\includegraphics[width=\linewidth,clip=]{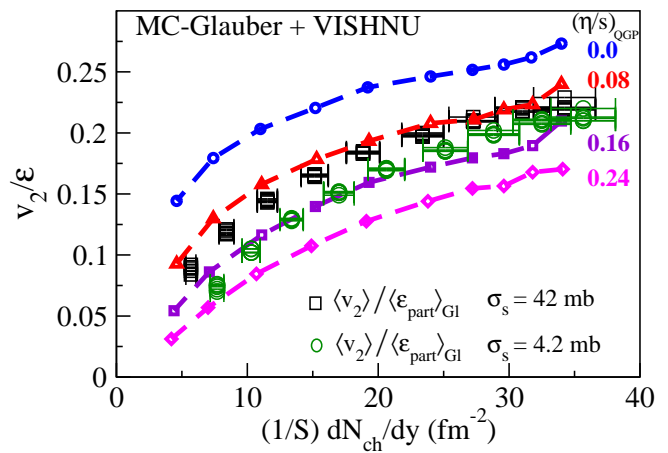}
\caption{\label{F9} (Color online)
Comparison of the universal theoretical curves for $v_2[\eta/s]/\varepsilon$ 
vs. $(1/S)(\dNdy)$ from the MC-Glauber model with $\sigma_s\eq42$\,mb
\cite{Song:2010mg} with experimental data for $\langle v_2\rangle$ 
\cite{Ollitrault:2009ie}, normalized by the eccentricity 
$\La\ecc_\mathrm{part}\Ra$ and transverse fireball area $S$ of 
the initial profile from the participant-plane averaged MC-Glauber model 
with standard ($\sigma_s\eq42$\,mb) and reduced ($\sigma_s\eq4.2$\,mb) 
smearing areas.
}
\end{figure}

Figure~\ref{F9} is a modified version of Fig.~2b in \cite{Song:2010mg}
which was used to extract the preferred value of $(\eta/s)_\mathrm{QGP}$
from experimental data by comparing them with \VC\ calculations using
participant-plane averaged initial conditions from the MC-Glauber model.
We remind the reader that, even though theoretically the dependence of
$v_2/\ecc$ on $(1/S)\dNdy$ is universal (at least at a fixed collision
energy \cite{Hirano:2010jg}) and depends only on a single parameter, 
$(\eta/s)_\mathrm{QGP}$, but not on the initial profile, we do not know 
the correct initial profile that drives the generation of elliptic flow 
in the actual experiments. Since different initial 
conditions have different eccentricities, the same set of experimental
$v_2$ and $\dNdy$ data yields different $v_2/\ecc$ and $(1/S)\dNdy$ when
normalized by $\ecc$ and $S$ from different initial state models, resulting
in different extracted values for $(\eta/s)_\mathrm{QGP}$ from a comparison 
with the universal theory curves. The larger eccentricities and smaller 
overlap areas resulting from a MC-Glauber initialization with reduced 
smearing area $\sigma_s\eq4.2$\,mb lead to smaller $v_2/\ecc$ and larger 
$(1/S)\dNdy$ values (green circled data in Fig.~\ref{F9}) than for the 
MC-Glauber model with standard smearing ($\sigma_s\eq42$\,mb, black squares). 
This results in a larger preferred value $(\eta/s)_\mathrm{QGP}$ (closer to
0.16 than the value of 0.08 we obtained when postulating MC-Glauber initial 
conditions with standard smearing). Furthermore, normalization of the 
experimental data with MC-Glauber ($\ecc,\,S$) values for reduced 
smearing changes the slope of the dependence of $v_2/\ecc$ 
on $(1/S)\dNdy$, and it no longer agrees with the slope of the universal
theoretical curves. We conclude that the comparison of experimental
data with \VC\ results disfavors the hypothesis that the experimentally
measured elliptic flow is generated by initial conditions that can be 
described by a MC-Glauber model with almost pointlike secondary particle
production. This conclusion aligns nicely with the theoretical prejudice
against such a model on the basis that it would violate the uncertainty 
principle, as discussed above. 

%
%


\end{document}